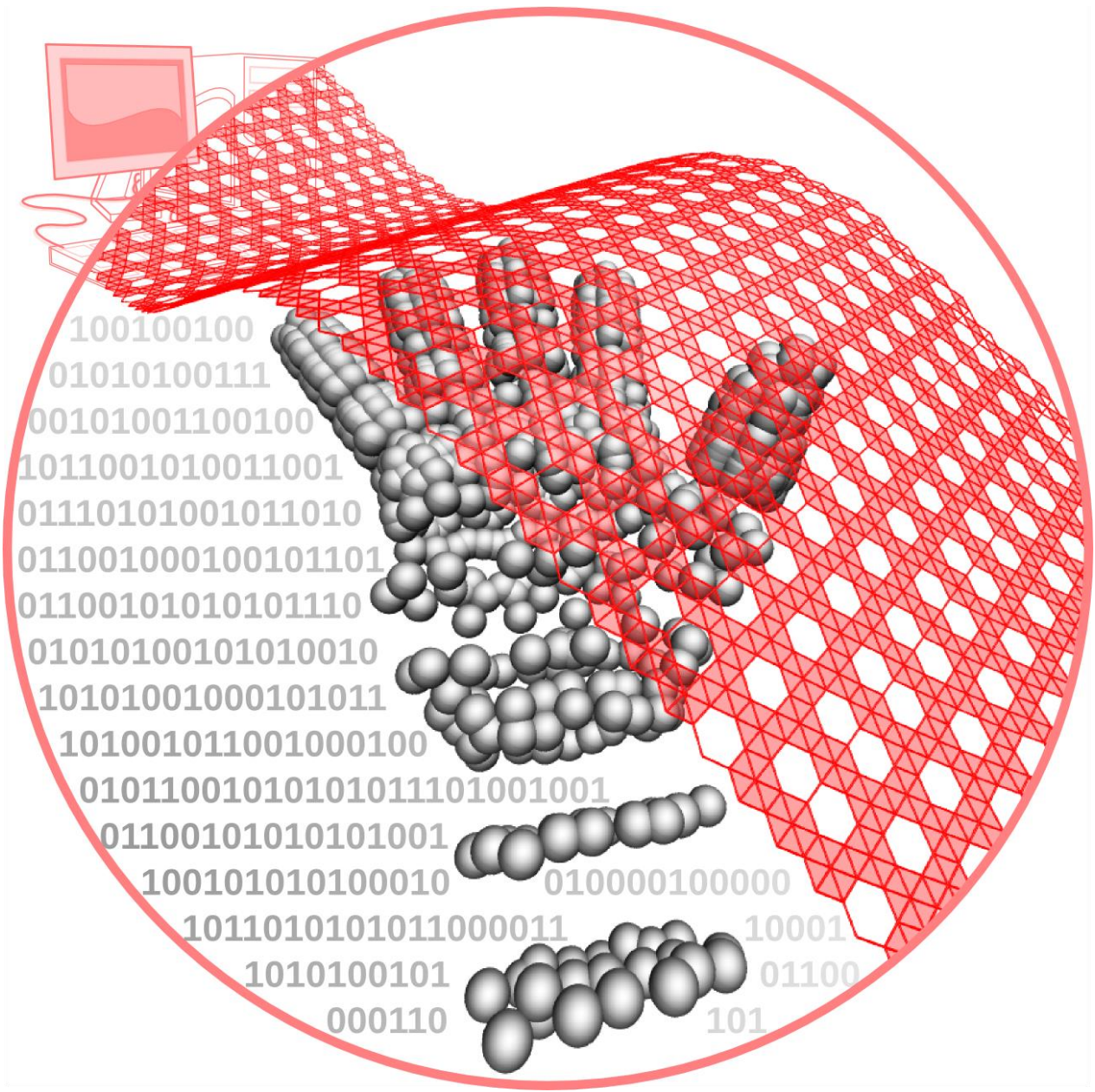

*Two-dimensional B Synthesis*

DOI:

# Probing Synthesis of Two-dimensional Boron by First-principles Computations

*Yuanyue Liu, Evgeni S. Penev, and Boris I. Yakobson\**

The synthesis of novel two-dimensional (2D) materials has attracted considerable interest due to their various unique properties. While the experimental realization of 2D boron ($^{2D}$B) sheets remains a challenge, it is important to theoretically investigate the possible fabrication methods. Here we explore the formation of B sheets on metal (Cu, Ag, Au) and metal boride ($MgB_2$, $TiB_2$) substrates via first-principles calculations. Our results suggest that B sheets can be grown on the Ag(111) or Au(111) surfaces by deposition. B atoms decomposed from precursor, and driven by the gradient of the chemical potential, will assemble into 2D clusters and further grow to a larger sheet, while formation of three-dimensional B ($^{3D}$B) structures could be impeded due to a high nucleation barrier. In addition, saturation of B-terminated $MgB_2$ surface in B rich environment can also lead to the formation of B sheets. These sheets are weakly bound to the substrates, suggesting feasible post-synthesis separation into the free-standing forms. Our work proposes feasible approaches to synthesize $^{2D}$B, and could possibly pave the way towards its application in nano-electronics.

The B sheets[1] are among those 2D materials which promise to have interesting properties and applications, yet remain impeded by the difficulties of experimental synthesis.[2] In particular, the metallic character of $^{2D}$B makes it a potential complement to graphene, h-BN, and metal-disulfides that may constitute ultimate building components in device designs. Unlike graphene, however, the accumulated theoretical knowledge on $^{2D}$B has not yet materialized experimentally. In contrast to previous theoretical studies,[1] focused on the structures and stability of hypothetical $^{2D}$B sheets, here we attempt to assess the feasibility of, and suggest some guidelines for practical routes towards their synthesis. Such theoretical investigation of the synthesis before its experimental realization is challenging, and cannot be comprehensive, but can quantify certain aspects and possibilities.

We propose and analyse synthesis methods, based on those established for graphene. One would be the exfoliation[3] from layered materials with weak inter-layer binding; however, no such precursor boron material exists in nature, which rules out the exfoliation approach. (i) *Deposition*, especially chemical vapor deposition[4] of B on catalytic substrate, either a pure metal or metal-terminated metal boride. The observed reconstruction of B-terminated boride surfaces[5] suggests another approach; (ii) *Saturation* by B deposition on the B-terminated surface of borides under B-rich environment, to form a B sheet; (iii) *Evaporation* of metal atoms[6] from metal borides at high temperature. In order to probe the feasibility of these methods, we perform first-principles calculations focusing on the following fundamental questions: (1) What are the atomic structures of $^{2D}$B on substrates and how they compare to the theoretically studied free-standing B sheets? (2) Is it energetically favorable to form $^{2D}$B on substrates, compared to other configurations such as B adatoms, clusters, or $^{3D}$B phases? (3) If formed, would it be possible to separate the B sheets from substrates after synthesis?

The choice of substrate is critical for the synthesis process. The epitaxial growth of graphene on Cu(111) surface[7] suggests that a good substrate should have low B solubility[8] while serving as a flat template for $^{2D}$B. Many metals are known to form borides, including those widely used for carbon nanotube growth (Fe, Co, Ni).[9] Group-11 elements, Cu, Ag, and Au, do not form borides, thus are selected here as substrates to model the deposition of B. In particular, the Ag(111) surface has been utilized in recent experiments on synthesizing 2D silicon[2d] and ZnO.[2a] The $^{2D}$B can also be expected to form from the borides which are composed of alternating metal and graphene-like B layers, e.g., $MgB_2$ and $TiB_2$, used here to represent the family of non-transition metal borides and the transition metal borides, respectively.

A broad variety of $^{2D}$B sheets can be described as $B_{1-x}V_x$ pseudoalloys,[11] where $V$ represents vacancy in a parent triangular lattice (for example, $x = 1/3$ corresponds to a hexagonal sheet). In vacuum, the energetically optimal fraction of vacancies falls in the range $0.1 < x < 0.15$, where numerous sheet structures with various vacancy patterns are found to be nearly degenerate in energy, suggesting a polymorphic state.[11] One of the most stable sheets, the α-sheet ($x = 1/9$),[20] and its structural relatives have been intensively studied.[1b, 1c, 1e, 1h-j, 1n, 1p, 1q] Here, ten types of B sheets with different vacancy concentration $x = v/27$ ($v = 0-9$), are constructed to explore general trends in properties as a function of $x$. Details of the construction procedure can be found in the Supporting Information (SI). These sheets are among the most stable ones in vacuum for given $x$.[11] With the above provisions, we can formulate more specifically the scenarios for synthesizing $^{2D}$B: (i) Deposition of B on the (111) surfaces of fcc Cu, Ag, Au, and the metal-terminated (0001) surfaces of $MgB_2$ and $TiB_2$; (ii) Saturation of B-terminated $MgB_2$ or $TiB_2$ surface in B rich environment; (iii) Evaporation of $MgB_2$ or $TiB_2$ at high temperatures. For each of these possibilities, we address the set of questions (1−3) posed above.

[*] Y. Liu, Dr. E. S. Penev, and Prof. B. I. Yakobson
Department of Mechanical Engineering and Materials Science, Department of Chemistry, and The Smalley Institute for Nanoscale Science and Technology, Rice University, Houston, Texas 77005, United States.
E-mail: biy@rice.edu

[**] This work was supported by the Department of Energy, BES Grant No. ER46598. The computations were performed at (1) the NICS, through allocation TG-DMR100029; (2) the NERSC, supported by the Office of Science of the DOE under Contract No. DE-AC02-05CH11231; and (3) the DAVINCI, funded by NSF under Grant No. OCI-0959097. We thank Zhuhua Zhang and Hoonkyung Lee for useful discussions.

Supporting information for this article is available on the WWW under http://www.angewandte.org or from the authors.

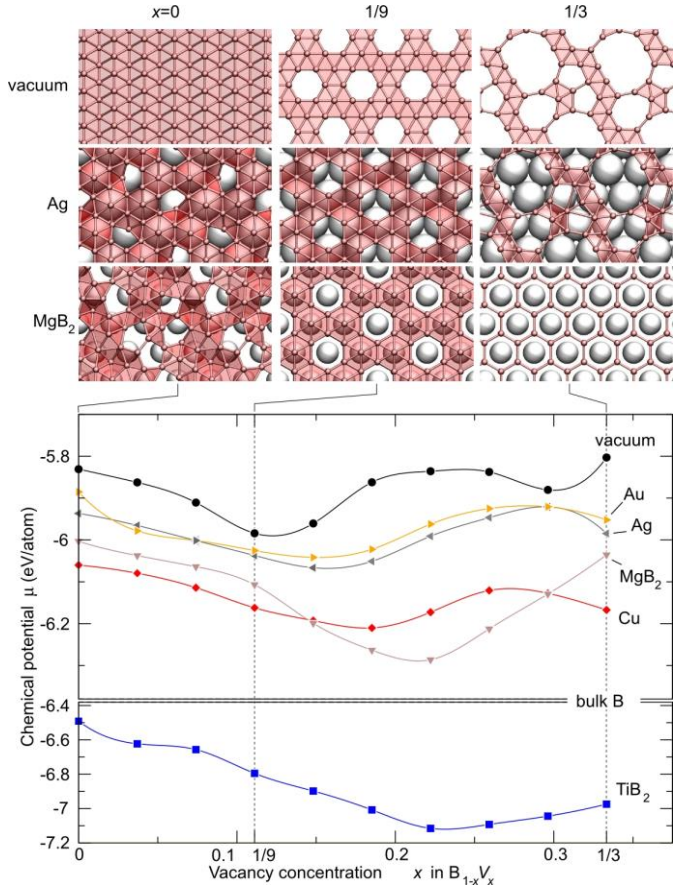

**Figure 1.** Atomic structure of $^{2D}$B sheets ($x$ = 0, 1/9, and 1/3) in vacuum and on substrates: Ag(111) and Mg-terminated MgB$_2$(0001) surfaces. $^{2D}$B is represented as a line network and the larger bright spheres are metal atoms (for clarity, only the topmost metal layer is rendered). The plot shows the chemical potential $\mu$ (Equation 1) of B sheets as a function of vacancy concentration $x$ in vacuum and on different substrates: Cu(111), Ag(111), Au(111), Mg-terminated MgB$_2$, and Ti-terminated TiB$_2$ surfaces. The vertical axis is broken at the $\mu$ of bulk α-B.

To compare the stability of different B sheets (the first question), we calculate the energy of B atoms in the corresponding configurations (essentially, the chemical potential $\mu$ at zero temperature, as shown in the SI) as:

$$\mu = (E_{\text{B-sheet/substrate}} - E_{\text{substrate}} - N_B E_{\text{B-atom}})/N_B \quad (1)$$

where $E_{\text{B-sheet/substrate}}$ is the total energy of the B sheet with the substrate, $E_{\text{substrate}}$ is the energy of the substrate without B, $E_{\text{B-atom}}$ is the energy of B atom in vacuum, and $N_B$ is the total number of B atoms in the system. The $\mu$ values of B sheets are shown in Figure 1.

In vacuum, we find that the free-standing hexagonal B sheet ($x$ = 1/3) cannot preserve its hexagonal lattice under small perturbations (in accord with the detected existence of unstable phonon modes). It would collapse into a disordered sheet with close-packed triangular domains and large voids, as shown in Figure 1. The nearly-amorphous sheet has energy much lower than the periodic hexagonal sheet by almost 0.7 eV/atom. In this case, although "vacancy concentration" is not a well-defined term, we can still use it for consistency with literature. In fact, at high vacancy concentration ($x > 5/27$), all the low-energy B sheets appear amorphous (structures are shown in the SI). The clustering of vacancies into large holes in B sheets has also been noted recently in other reports[1c, 10]. The amorphous structures are also known to exist in the B fullerenes.[11] Nevertheless, at low $x < 6/27$, the hexagonal framework is sustained and the α-sheet is the most stable one in this set.

When placed on Cu, Ag, Au, the B sheets with $x > 6/27$ are also amorphous. Figure 1 shows the atomic structures of B sheets on Ag. At low B coverage (high $x$), the sheet is composed of stripes and voids, while it gradually restores the hexagonal framework at high B coverage (low $x$). In contrast, on MgB$_2$ and TiB$_2$, the B sheets with high $x$ retain the hexagonal lattice, as a natural continuation of the bulk structure. One common feature shared by the B sheets is that they become more buckled with the decrease of $x$ (increase of B coverage). In vacuum, the amorphous sheet is flat, while the α-sheet buckles by 0.4 Å,[1e] and the triangular sheet has height variation of 0.9 Å. On Ag, the buckling increases from 0.4 Å for $x$ = 1/3, to 0.8 Å for $x$ = 1/9, then to 1.8 Å for $x$ = 0. Similarly, on MgB$_2$, it varies from flat, to 0.4 Å, and then to 1.6 Å. Figure 1 also shows that on substrates the α-sheet is no longer the most stable one. Interestingly, the optimal vacancy concentration of $^{2D}$B sheets shifts to a higher $x$ value, that is lower density, compared to that in vacuum.

Which type of B sheet could be synthesized depends on the formation route: (i) In deposition, B adatoms from the decomposed precursor should possibly assemble into a sheet with the lowest $\mu$ in Figure 1. For illustration, Figure 2 shows the atomic structures on MgB$_2$ and Ag. On the Mg-terminated surface of MgB$_2$, the B sheet has a hexagonal framework located on top of 3-fold hollow sites at ~1.8 Å distance from the surface, which is slightly larger than the distance between the B layer and Mg layer in bulk MgB$_2$ ~1.7 Å. A third of the hexagon centers ($x$ = 2/9) are filled with B atoms, which sit on top of metal atoms with ~0.4 Å separation from the hexagonal plane. On Ag, the B sheet (β-type[1e]) is flatter ($x$ = 4/27) but more distant from the substrate, ~2.5 Å, implying weaker interaction.

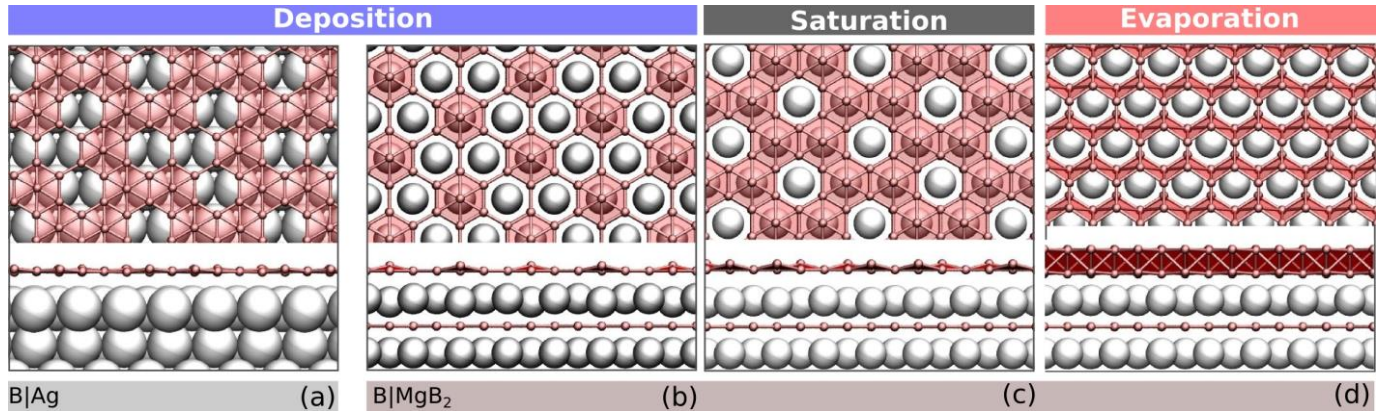

*Figure 2.* Atomic structure of B sheets, representing different synthesis methods: (a) deposition on metals and (b) metal-terminated surface of borides, (c) saturation of B-terminated surface of borides, and (d) evaporation of borides. Ag and $MgB_2$ are shown as representatives for metals and metal borides, respectively. Both top and side views are shown.

Unlike the deposition method, where the lowest-$\mu$ sheet is expected to grow,[12] methods (ii) and (iii) lead to different forms of $^{2D}B$. (ii) For surface saturation, B atoms from the environment fill in the centers of B hexagons on the B-terminated surface.[5] The vacancy concentration is fully determined by $\mu$ of B in the environment. By tuning $\mu$, one can in principle obtain various B sheets, from hexagonal to triangular,[13] in contrast to the mono-type sheet produced by method (i). Figure 2c shows the sheet with the weakest adhesion to $MgB_2$ substrate ($x = 1/9$, i.e., the α-sheet, at ~2.2 Å distance from the surface). (iii) At the surface of metal borides, two hexagonal B layers are intercalated with one metal layer. According to the phase diagram of $MgB_2$,[14] at ~1100 K and 1 atm pressure, the Mg atoms sublimate, leaving behind two hexagonal B layers inter-connected and forming a hexagonal bilayer on the surface, Figure 2d. Our calculations show that the interlayer distance of the bilayer is only 1.5 Å. A similar mechanism operates in the formation of graphene *via* SiC evaporation,[15] where, in contrast, the two carbon hexagonal layers bind with each other very weakly,[16] resulting in an easy separation of the top graphene sheet from the underlying material.

In methods (ii)-saturation and (iii)-evaporation, hexagonal B layers pre-exist on the substrate. However, in method (i)-deposition, it is not obvious whether the B adatoms on the substrate would aggregate, or would tend to form disconnected populations. We find that on all the substrates, $\mu_{adatoms} > \mu_{2D-cluster} > \mu_{sheet}$, providing a driving force for growth. Figure S3 in the SI shows the decrease of $\mu$ as B assembles on Ag as an example, which asymptotically approaches $\mu$ of B sheet. However, as shown in Figure 1, aside from $TiB_2$, the other substrates cannot decrease $\mu$ of $^{2D}B$ below the value for bulk $^{3D}B$.[17] At first glance, the bulk $^{3D}B$ instead of the 2D sheet could be the growth product on these substrates. However, as shown in the SI, the high nucleation energy of $^{3D}B$ could impede its formation, and thus the small B nuclei always have planar structures.[18] In addition, the low solubility of B into substrates and fast diffusion on surfaces could help the B adatoms transport in a 2D channel,[8] and grow the nucleus into a monolayer structure. The detailed analysis can be found in the SI.

The question (3) posed in the beginning can be addressed by calculating the adhesion energy, the work required to separate $^{2D}B$ from the substrate. Figure 3 shows the adhesion energies for different growth methods. As discussed above, growth by surface saturation of borides can result in various 2D-forms depending on $\mu$ of B in the environment, but only weakly bound ones can be separated from substrate. Therefore the sheet with lowest adhesion energy is chosen for comparison. The adhesion energies of graphite[19] and graphene on Cu(111) surface[20] are taken from the literature. The $TiB_2$ surface has strong adhesion with the B sheet, > 0.2 eV/Å$^2$, in agreement with the high cohesive energy of bulk $TiB_2$.[9b] Only the $^{2D}B$ grown by deposition on Ag, Au, or saturation of B-terminated $MgB_2$ surface has adhesion energy of the same order of magnitude as graphene on Cu, < 0.10 eV/Å$^2$, thereby narrowing the synthesis options to deposition on Ag, Au, and saturation of B-terminated $MgB_2$ surface. The weak adhesion of the B sheet to Ag and Au surfaces has already been hinted at above by the large distance from the substrates ~2.5 Å. The charge density difference plots shown in the inset of Figure 3 illustrate how the electrons are redistributed upon the sheet-substrate contact formation. The electrons are transferred from Ag to B, while the Mg atoms under B sheet retain their original ionic states. Both substrates cause the depletion of in-plane σ states and accumulation of out-of-plane π electrons. However, it should be noted that the computational method employed here (density-functional theory, PBE exchange-correlation functional) generally underestimates the adhesion energy between substrates and adsorbates,[20] and an efficient computational scheme to obtain very accurate values is still lacking. Nevertheless, the current method is helpful, at least, to screen out the strongly-adhesive substrates.

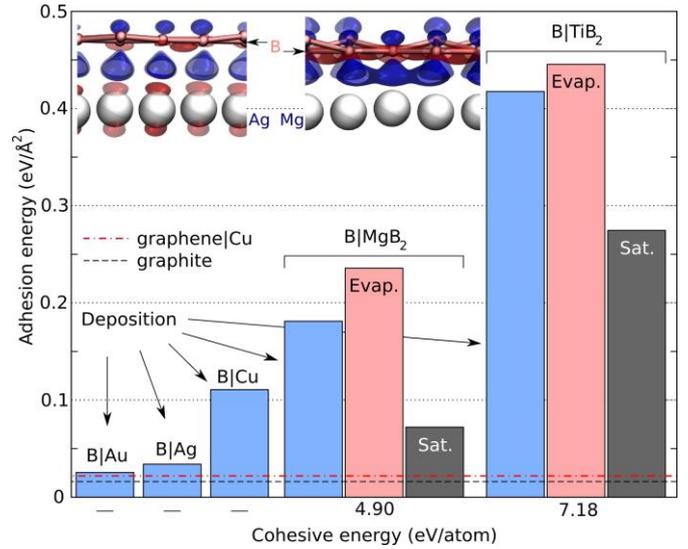

*Figure 3.* Adhesion energies of B sheets to substrates, representing different synthesis methods: deposition on Cu, Ag, Au, $MgB_2$ and $TiB_2$; evaporation of borides, saturation of B-terminated surfaces of borides-$MgB_2$ and $TiB_2$. The adhesion energies of graphite[19] and graphene on Cu[20] are indicated. The cohesive energy values are those of bulk $MgB_2$ and $TiB_2$. The inset shows the charge density difference isosurfaces for B sheet on Ag and $MgB_2$. The electrons accumulation is shown in blue and depletion is in red. The B layers are represented as light red ball-and-sticks and the topmost metal layers are shown as gray circles.

In conclusion, our analysis of growth scenarios based on first-principles calculations suggests the possibility of synthesizing $^{2D}B$ on metals or metal boride surfaces. (i) Deposition of B on Au or Ag(111) surface can result in growth of $^{2D}B$, whereas a high nucleation barrier could impede the formation of 3D-structures; (ii) Saturation of B-terminated $MgB_2$ surface under B-rich conditions can also lead to the formation of $^{2D}B$. In both cases, the B sheets bind weakly to the substrates, suggesting a relatively easy post-growth separation. The present analysis may be potentially beneficial not only for B but could also provide a roadmap for theoretical investigation of growth of 2D-materials in general, and possibly help with their experimental realizations.

**Keywords:** boron · density functional calculations · synthesis design

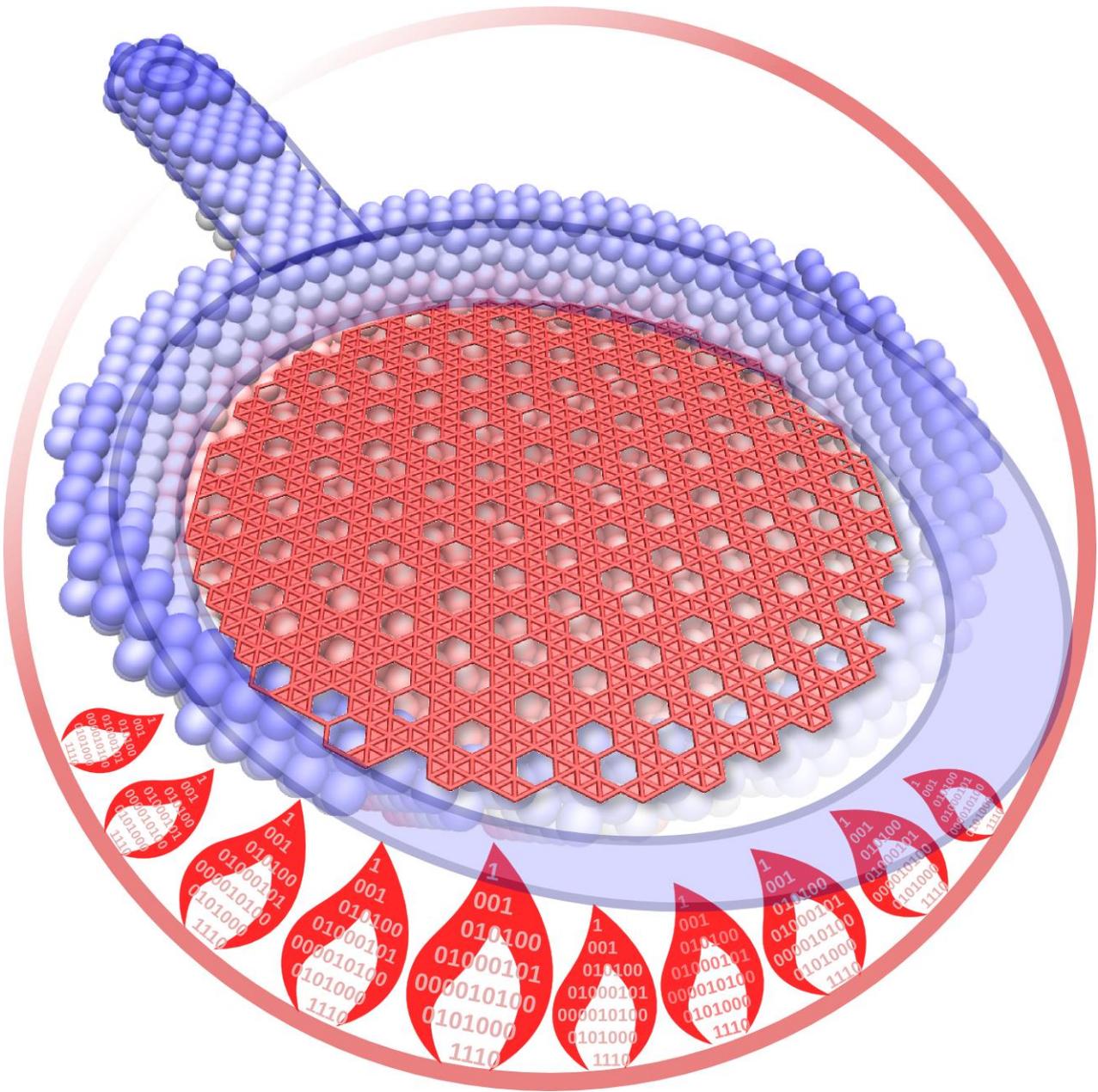